\documentclass[12pt]{article}

\usepackage{scicite}

\usepackage{times}
\usepackage{lscape}
\usepackage{graphicx}

\newenvironment{sciabstract}{%
\begin{quote} \bf}
{\end{quote}}

\newcounter{lastnote}
\newenvironment{scilastnote}{%
\setcounter{lastnote}{\value{enumiv}}%
\addtocounter{lastnote}{+1}%
\begin{list}%
{\arabic{lastnote}.}
{\setlength{\leftmargin}{.22in}}
{\setlength{\labelsep}{.5em}}}
{\end{list}}

\title{Early Optical Polarization\\ 
of a\\
Gamma Ray Burst Afterglow}

\author
{Carole G. Mundell,$^{1\ast}$ Iain A. Steele,$^{1}$ Robert
J. Smith,$^{1}$ Shiho Kobayashi,$^{1}$\\Andrea Melandri,$^{1}$
 Cristiano Guidorzi,$^{1,2,3}$ Andreja Gomboc,$^{4}$\\ Chris J. Mottram,$^{1}$
  David Clarke,$^{5}$ Alessandro Monfardini,$^{1,6}$\\ David Carter,$^{1}$
  David Bersier$^{1}$\\
\\
\normalsize{$^{1}$Astrophysics Research Institute, Liverpool John Moores University,}\\
\normalsize{Twelve Quays House, Egerton Wharf, Birkenhead, CH41 1LD, U.K.}\\
\normalsize{$^{2}$Dipartimento di Fisica, Universit\'a di Milano-Bicocca, Piazza delle Scienze 3,}\\
\normalsize{20126 Milano, Italy}\\
\normalsize{$^{3}$INAF-Osservatorio Astronomico di Brera, via Bianchi 46, 23807 Merate (LC), Italy}\\
\normalsize{$^{4}$Faculty of Mathematics and Physics, University of Ljubljana, Jadranska 19,}\\
\normalsize{1000 Ljubljana, Slovenia}\\
\normalsize{$^{5}$Department of Physics \& Astronomy, University of Glasgow, G12 8QQ, UK}\\
\normalsize{$^{6}$CNRS, Institut N\'eel, 25 Avenue des Martyrs, 38042 Grenoble, France}\\
\\
\normalsize{$^\ast$To whom correspondence should be addressed; E-mail:  cgm@astro.livjm.ac.uk.}
}

\date{}

\begin{document} 




\def\lesssim{\mathrel{\hbox{\rlap{\hbox{\lower4pt\hbox{$\sim$}}}\hbox{$<$}}}}
\def\moresim{\mathrel{\hbox{\rlap{\hbox{\lower4pt\hbox{$\sim$}}}\hbox{$>$}}}}
\maketitle 


\begin{sciabstract} 

We report the optical polarization of a gamma ray burst (GRB) afterglow,
obtained 203~seconds after the initial burst of $\gamma$ rays from
GRB~060418, using a ring polarimeter on the robotic Liverpool
Telescope.  Our robust (2-$\sigma$) upper limit on the percentage of
polarization, less than 8\%, coincides with the fireball deceleration time at
the onset of the afterglow. The combination of the rate of decay of
the optical brightness and the low polarization at this critical time
constrains standard models of GRB ejecta, ruling out the presence of a
large-scale ordered magnetic field in the emitting region.

\end{sciabstract}



\noindent Gamma ray bursts (GRBs) are the most instantaneously
powerful explosions in the Universe and represent the most important
new astrophysical phenomenon since the discovery of quasars and
pulsars. Identified as brief, intense and unpredictable flashes of
high-energy $\gamma$ rays on the sky, the most common type of
GRB, so-called long bursts, have $\gamma$-ray pulses that last longer
than 2~s. These are thought to be produced when a massive star reaches
the end of its life, its core collapsing to form a black hole and, in
the process, ejecting an ultra-relativistic blastwave
\cite{piran04,mes06}. In many cases, the detected $\gamma$-ray flux
implies an unphysically high explosion energy if assumed to be emitted
isotropically by the source, the so-called energy
catastrophe. Instead, focussing the energy into a narrow jet reduces
the intrinsic energy output to a canonical $\sim 10^{51}$~erg for most
GRBs \cite{frail01}.

After the initial burst of $\gamma$ rays, the subsequent radiation
produced at longer wavelengths (e.g x-ray, optical or radio), termed the
"afterglow", is generally accepted to be synchrotron radiation whose
observed properties are consistent with a focussed jet expanding at
ultra relativistic speeds into the interstellar medium.  The
production of synchrotron radiation requires the presence of a
magnetic field but the origin and role of the magnetic fields in GRB
ejecta are a long-standing open issue.  In turn, fundamental questions
on the driving mechanism of the explosion, in particular, whether the
relativistic outflow is dominated by kinetic (baryonic) or magnetic
(Poynting flux) energy, remain unanswered \cite{piran05,lyu06}.  The
primary challenges in addressing these issues arise because GRBs are
short-lived, compact and lie at vast cosmological distances; our
understanding of their physical nature is therefore inferred from the
characteristics of their radiation, measured at the earliest possible
time when the observed radiation is still sensitive to the properties
of the original fireball.

The two main models of collimated relativistic outflows, or jets, that
have been proposed are the hydrodynamical and the magnetized jet
\cite{lyu06}. Hydrodynamical jets have no dominant ordered magnetic
field but instead produce synchrotron radiation from tangled magnetic
fields, concentrated in the thin layer of the expanding shock front,
that are generated locally by instabilities in the shock
\cite{gruwax99}; the magnetic field does not influence the subsequent
evolution of the jet. Models of these jets have been highly successful
at reproducing a wide range of observed properties of GRBs
\cite{piran04,mes06}. A relativistic outflow from a central engine
might have a weak ordered or random magnetic field. As long as
the magnetic field does not affect the dynamics of the jet, we
classify it as a hydrodynamical jet.  In contrast, magnetized jets are
threaded with strong, globally ordered magnetic fields, which
originate at the central source, are advected outward with the
expanding flow and may provide a powerful mechanism for collimating
and accelerating the relativistic jet \cite{drenk02,lyu03}.  A
magnetic driving mechanism is an attractive scenario to account for
the prodigious energy outputs and vast accelerations required for GRB
ejecta, as well as for over-coming energy efficiency problems inherent
in hydrodynamical models in which internal shocks must convert kinetic
energy to radiative energy with sufficient efficiency to produce the
observed $\gamma$ ray emission and prolonged central engine activity
\cite{nousek06,zhang06}.

Observationally, the fading rate of the afterglow emission alone is
inadequate as a diagnostic for distinguishing between these
theoretical jet models \cite{granot03,rossi04,laz04}; in contrast, the
polarization properties are predicted to differ markedly.
Observations of the polarization state of GRB afterglow emission
therefore offer a  diagnostic to eliminate or constrain
current models. The testable prediction is that hydrodynamical jets
produce a considerable amount of polarization at the geometrical transition
phase a few days after the burst, the so-called 'jet break' time when
the lateral spreading of the slowing jet produces a characteristic
steepening of the light curve, and produce little or no polarization at
early times, whereas jets with large-scale globally-ordered magnetic
fields produce polarization substantially greater than 10\% at early
times \cite{rossi04,laz04} and in some cases as high as $\sim$50\%
\cite{laz04}.

The first detection of polarized optical emission from a GRB afterglow
was taken at 0.77 days after the burst of GRB~990510 and, with the
exception of GRB~020405 for which an unexplained high degree of
optical polarization was measured 1.3 days after the burst
\cite{bersier03}, late-time measurements of optical polarization for
other long bursts taken typically at $t\moresim$ 0.2 day all show
consistently low values of P$\sim$1 to 3\%, some of which may be induced by
interstellar scattering processes
\cite{covino99,covino02,greiner03,barth03}. Although these painstaking
observations of late-time polarization were vital in confirming the
presence of collimated jets in GRBs [e.g. \cite{covino99,covino02,greiner03}], there was a lack of
polarization observations of GRB afterglows in the early phase within
the first few minutes, where the predicted properties of magnetized or
unmagnetized hydrodynamic jets differ most.

Recent advances in technical efficiency of catching the rapidly fading
light from GRBs, driven primarily by the real-time dissemination of
accurate localizations of GRBs discovered by the Swift satellite
\cite{Gehrels04}, have opened a new era in rapid-response followup
studies of GRBs and their afterglows \cite{piran04,mes06}.

GRB~060418 was detected by the Swift satellite at 03:06:08 UT on 18
April 2006 and exhibited a triple-peaked $\gamma$-ray light curve with
overall duration of $\sim$52 s, followed by a small bump at 130
seconds coincident with a large flare detected in the x-ray light
curve and likely associated with ongoing central engine activity
\cite{gcn5009}. A localization was communicated automatically to
ground-based facilities, and triggered robotic followup observations
at the 2.0-m Liverpool Telescope in La Palma, the Canary Islands.
These observations consisted of a 30-s exposure with the RINGO
polarimeter (Fig. 1) beginning at 03:09:31 UT or 203 s after the start
of the prompt $\gamma$-ray emission and contemporaneous with the
fading tail of this $\gamma$-ray emission, followed by 2 hours of
multi-color photometric imaging. We concentrate here on the RINGO
measurement.

RINGO uses a rotating polaroid to modulate the incoming beam, followed
by co-rotating deviating optics that transform each star image into a
ring that is recorded on the charge-coupled device (CCD) chip. Any
polarization signal present in the incoming light is mapped out around
the ring in a sin~($2\theta$) pattern. A description of the instrument
and the data reduction procedures are given in \cite{supp1}.  A bright
star in the field of view of the GRB (Fig. 1) was used as a check on
our data reduction, with multiple measurements made on subsequent
nights confirming its measured polarization in the GRB frame of
$<1\%$.  This value also provides a lower limit to any contribution of
polarization that could have been induced into the GRB by galactic
interstellar dust.  No appreciable polarization signal could be
detected from the GRB.  To quantify this, we carried out a Monte Carlo
error analysis was carried out in an attempt to recover
an artificially induced polarization signal with a noise spectrum
identical to that of the GRB data.  This gave a firm (2$\sigma$) upper
limit to the measured polarization of $<8$\% [polarizations of 10\%
for example being easily detectable - see \cite{supp1}].

The optical and near infrared light curves of GRB~060418 are smooth
and featureless; the IR light curves show a smooth rise ($\alpha \sim
2.7$ where $F \propto t^{\alpha}$) to a broad peak at time t$_{peak}
\sim 153$~s \cite{molin06} before fading away with a smooth, unbroken
power law with $\alpha \sim -1.2$, identical to the decay
rate of the optical light curves and typical of standard fireball
models of optical afterglows.  In the standard GRB fireball model in
which a jet is driven into the surrounding circumburst medium, the
early afterglow light is thought to include contributions from both a
forward shock, which propagates into the ambient medium, and a reverse
shock, which propagates back into the original fireball ejecta
\cite{kps99}.  Forward-shock emission
peaks when the fireball decelerates or when the typical synchrotron
frequency ($\nu_m$) passes through the observed band.  The lack of
colour change around the peak in the IR light curves of GRB~060418
\cite{molin06} confirms the deceleration interpretation, with $\nu_m$
already lying below the optical and IR bands at this time
\cite{molin06}.  The steep temporal rise of the IR light curve
($\alpha \sim 2.7$) is also consistent with theoretical predictions
of forward shock emission before deceleration \cite{kz03}.

The RINGO measurement was made close to the time of the peak of the IR
light curve at the fireball deceleration time and onset of the
afterglow, making the polarization measurement particularly important
for testing afterglow predictions from current standard jet models.
Our polarization measurement also coincides with the decay phase of
the x-ray flare emission.  Extrapolating the peak flux density in the
x-ray flare at 130~s to optical wavelengths and assuming a spectral
index between optical and x-ray bands of $\beta \sim 1$ ($F_{\nu}
\propto \nu^{-\beta}$), we found that the maximum contribution of the
flare to the optical band is negligible, thus ruling out an internal
shock origin for the optical emission and confirming that the optical
emission represents the afterglow at the time of the RINGO
measurement.

Although the optical emission from GRB~060418 was bright at early
time, no dominant optical flash from the reverse shock was detected,
similar to other recently studied bright bursts such as GRB~061007
\cite{mundell07}. The apparent lack of an optical or IR flash is
easily explained in the standard fireball model if the typical
synchrotron frequency of the forward shock emission, $\nu_m$, is lower
than the observing frequency of the optical (and IR) band,
$\nu_{opt}$, at the onset of afterglow, or the peak time
$t_{peak}$. This condition is required also to interpret the IR light
curve peak, otherwise the rise gradient is expected to be shallower,
$t^{1/2}$, than observed \cite{spn98}. Mundell et al. \cite{mundell07}
suggested that a low value of $\nu_m$ may be produced by small
microphysics parameters, in particular low $\epsilon_B$, due to small
magnetic fields in forward and reverse shock regions. A low typical
synchrotron frequency can also result if the fireball is enriched with
electron-positron pairs. Non-standard models of hydrodynamical jets
with weak magnetic fields that radiate via inverse Compton emission,
rather than synchrotron emission, have also been proposed as a
mechanism for suppressing optical flashes.  No polarization
predictions for non-standard models exist so we do not discuss these
models further.  Instead, we test predictions from standard GRB models
of relativistic jets with and without globally ordered magnetic fields
that emit synchrotron radiation.

Theoretical models of magnetized jets, with large-scale ordered
magnetic fields originating from the central engine, predict high
values of polarization at very early times for the prompt $\gamma$-ray
emission \cite{rossi04,laz04}.  Putative detections of large levels of
$\gamma$-ray polarization of $\sim$70\% to 80\% \cite{cb03} and
$>$35\% and $>$50\% \cite{willis06} in a small number of GRBs provide
support for large-scale ordered magnetic fields in the region of the
flow that produces the high-energy prompt emission, but the
observational results remain controversial \cite{rf04}.  The optical
emission from the forward shock is also predicted to be highly
polarized for these magnetized jets; instabilities in the contact
discontinuity at the fireball surface are expected to act as anchors
for continuing the ordered magnetic field into the afterglow emission,
producing optical polarization as high as 10\% to  50\% at early time
\cite{granot03,laz04,rossi04,swl04}. The exact level of observed
polarization depends on complex details of the degree of mixing
between the ordered magnetic field in the ejecta and any tangled
component in the shock front. Nevertheless, the key characteristic of
emission from jets with large-scale ordered magnetic fields is that
the observed polarization does not disappear at very early times
\cite{laz04}.

Our robust upper limit $P<8\%$ at the very early time t$\sim$203~s for
GRB~060418 lies below predicted values for reasonable jet properties.
In the standard synchrotron shock model, the temporal decay rate of
the optical afterglow, $\alpha$, is related to the underlying power
law distribution of electron energies, or $dn/de = e^{-p}$; for
GRB~060418, we derive $p=2.6$, typical of optical afterglow
emission. Theoretical models of a magnetically dominated flow for
$p=2.6$ predict observable polarization of a few tens of percent
\cite{lyu03}, substantially larger than that observed for
GRB~060418. Within the limitations of current theoretical models the
low level of polarization observed in GRB~060418 therefore indicates
that large-scale ordered magnetic fields are not dominant in the
afterglow emission at early times.

Although reverse shock emission in the form of an optical flash does
not dominate the light curves of GRB~060418, in the hydrodynamic
jet model more than $\sim$50\% of the emitted photons come from the
original fireball material \cite{mundell07}, or reverse shock region,
at the deceleration time when our polarization measurement of
GRB~060418 was made.  This is because at the peak time, the two shock
emissions have the same cooling frequency, and the peak values of $\nu
F_\nu$ at the cooling frequency are comparable. The two emissions
contribute equally to the total flux at observing frequencies between
the cooling frequency and the typical frequency of the forward shock
\cite{kz03}, as is the case for optical measurements of GRB~060418.

We therefore rule out the presence of a magnetic field with ordered
large-scale structure in a hydrodynamic or baryonic jet, in which the
energy density of any magnetic field component is comparable to or
less than that of the baryonic component, because this would also
result in a large amount of polarization at early time.

Our result is consistent with the theoretical prediction of low or
zero polarization for hydrodynamical jets
 without large-scale ordered magnetic fields when observed at early
 times \cite{laz04}. This is also consistent with the reported lack of
 linear or circular polarization at radio frequencies for the
 afterglow of GRB~991216, observed at t$\sim$1~day after the burst
 \cite{gt05}.  Thus we support models of hydrodynamical jets in which
 the generation of the magnetic field in the regions responsible for
 both the prompt and afterglow emission is driven by local processes
 in the fluid.

{}

\begin{itemize}

\item Supporting Online Material:\\ 
www.sciencemag.org/cgi/content/full/1138484/DC1 

\item Materials and Methods \\
Figs. S1 to S4 

\item References and Notes\\ 
6 December 2006; accepted 2 March 2007\\ 
Published online 15 March 2007; 10.1126/science.1138484\\ 
Include this information when citing this paper.\\ 

\end{itemize}
\clearpage

\clearpage

\begin{figure}
\hspace{-1.3cm}
\includegraphics[width=140mm,angle=270]{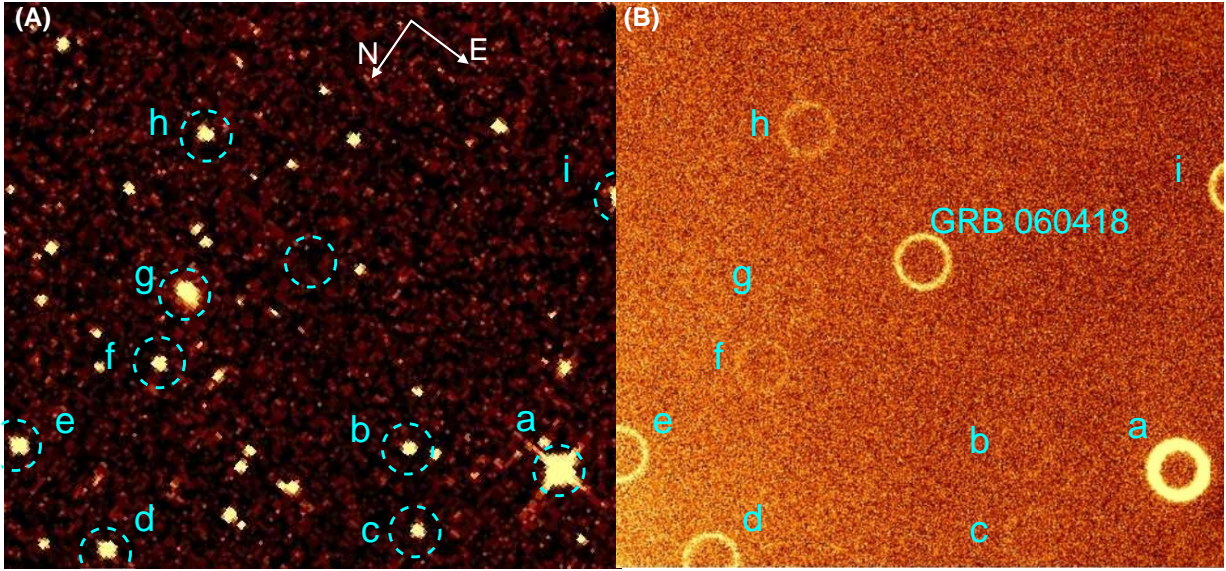}
\caption{Direct optical and RINGO polarimeter images of
the field containing GRB~060418. The direct R-band image (A) is taken
from the Digital Sky Survey (DSS) and shows the sky before the GRB
occurred.  The RINGO image (B) consist of a CCD recording of the
incoming light from GRB~060418 and other bright sources in the field
after the light has been modulated by a rotating polaroid and spread around
rings by co-rotating deviating optics. The objects detected by RINGO
are labelled (a) to (i) in both panels and blue dotted rings,
corresponding to those in the RINGO image, are shown on the DSS image
as a guide. All labelled objects, with the exception of extended
object (g), are unresolved point sources and thus produce well-defined
rings. The bright star (a) was used for additional calibration as
described in the text. The field of view is 4.6 by 4.6 arc min and
the orientation of the field is shown by the white arrows indicating
north (N) and east (E).}
\end{figure}

\end{document}